\documentclass{PoS}

\title{CTEQ nuclear parton distribution functions}

\ShortTitle{Short Title for header}

\author{\speaker{K. Kova\v{r}\'\i k}\\
        Institut f\"ur Theoretische Physik, Universit\"at Karlsruhe, KIT, 76128 Karlsruhe, Germany\\	
        E-mail: \email{karol.kovarik@kit.edu}}
\author{T. Je\v{z}o\\
        Department of Mathematical Sciences, University of Liverpool, Liverpool L69 3BX, UK\\
        E-mail: \email{T.Jezo@liverpool.ac.uk}}
\author{A. Kusina\\
        Southern Methodist University, Dallas, TX 75275, USA\\
        E-mail: \email{akusina@smu.edu}}
\author{F. I. Olness\\
        Southern Methodist University, Dallas, TX 75275, USA\\
        E-mail: \email{olness@smu.edu}}
\author{I. Schienbein\\
        Laboratoire de Physique Subatomique et de Cosmologie, Universit\'e Joseph Fourier/CNRS-IN2P3/INPG, 
		53 Avenue des Martyrs, 38026 Grenoble, France\\
        E-mail: \email{schien@lpsc.in2p3.fr}}
\author{T. Stavreva\\
        Laboratoire de Physique Subatomique et de Cosmologie, Universit\'e Joseph Fourier/CNRS-IN2P3/INPG, 
		53 Avenue des Martyrs, 38026 Grenoble, France\\
        E-mail: \email{stavreva@lpsc.in2p3.fr}}
\author{J. Y. Yu\\
        Southern Methodist University, Dallas, TX 75275, USA\\
        E-mail: \email{yu@physics.smu.edu}}

\abstract{We show for the first time preliminary results of nuclear parton distribution function analysis of charged lepton DIS and 
Drell-Yan data within the CTEQ framework
including error PDFs. We compare our error estimates to estimates of different nPDF groups.
}

\FullConference{XXI International Workshop on Deep-Inelastic Scattering and Related Subject -DIS2013,\\
		22-26 April 2013\\
		Marseilles,France}

\begin{document}
\section{Introduction}
At an era of hadron colliders such as Tevatron and LHC, every prediction tested there requires parton distribution functions (PDFs)
which describe the structure of colliding hadrons in terms of quarks and gluons. 
The importance of PDFs is why many groups perform and update global analyses of PDFs for protons \cite{Ball:2009mk, 
Martin:2009iq, Nadolsky:2008zw,Owens:2012bv} and for nuclei \cite{Hirai:2007sx,Eskola:2009uj,deFlorian:2011fp}.

Nuclear effects in parton distribution functions are important not only in predictions involving nuclei directly
but also parton distribution functions of protons are fitted using also fixed target experiments taken on nuclear targets, 
mainly deuterium  but also heavy nuclei such as lead and iron in case of neutrino DIS. 
Moreover experiments involving heavy ions at RHIC and LHC require dedicated parton distribution functions which 
include nuclear effects systematically - nuclear PDFs (nPDFs). The nuclear effects in nPDFs are typically added on top of 
existing proton PDFs and are independently fitted from experimental data on nuclei.

Although in theory PDFs of free protons and nPDFs of protons bound in nuclei are derived from the same basic principles 
such as factorization and perturbative QCD and both analyses include predominantly theory predictions at the
next-to-leading order, the current state of the art of the two analyses is much different. 
Proton PDF fits use a large and very precise data sample from HERA and Tevatron where as nuclear PDFs 
are fitted to a variety of smaller nuclear data samples from several fixed target experiments 
taken on different nuclei and some collider data from RHIC. The amount and precision of the nuclear data is far
inferior to the data available for free protons. On top of that nuclear PDFs have the ambition to describe
parton distributions for each nucleus and so there are more free parameters in a typical nPDF fit compared to 
a free proton analysis. As a result lacking precision and more free parameters the uncertainty of nuclear parton distribution 
functions is much larger than that for the free protons. 
Therefore it is imperative to compare different error analysis of different nPDF in order to correctly estimate the true uncertainty.
\section{Nuclear PDFs in the CTEQ framework.}
Here we present an updated nuclear PDF analysis originally presented in \cite{Schienbein:2009kk} and \cite{Kovarik:2010uv} 
where the parameterizations of the nuclear parton distributions of partons in bound protons at the input scale of $Q_0=1.3$ GeV are	
\begin{eqnarray}
x\, f_{k}(x,Q_{0}) &=& c_{0}x^{c_{1}}(1-x)^{c_{2}}e^{c_{3}x}(1+e^{c_{4}}x)^{c_{5}}\,,\\ \nonumber
\bar{d}(x,Q_{0})/\bar{u}(x,Q_{0}) &=& c_{0}x^{c_{1}}(1-x)^{c_{2}}+(1+c_{3}x)(1-x)^{c_{4}}\,,
\end{eqnarray}
where $f_k=u_{v},d_{v},g,\bar{u}+\bar{d},s,\bar{s}$ and $\bar{u},\bar{d}$
are a generalization of the parton parameterizations in free protons used in the CTEQ proton analysis \cite{Pumplin:2002vw}. 
To account for different nuclear targets, the coefficients $c_k$ are made to be functions of the nucleon number $A$
\begin{equation}
c_{k}\to c_{k}(A)\equiv c_{k,0}+c_{k,1}\left(1-A^{-c_{k,2}}\right),\ k=\{1,\ldots,5\}\,.\label{eq:Adep}
\end{equation}
In the current analysis, the same standard kinematic cuts $Q>2\ {\rm GeV}$ and $W>3.5\ {\rm GeV}$ were applied as in 
\cite{Pumplin:2002vw} and we obtain a fit with $\chi^{2}/{\rm dof}$ of 0.87 to 708 data points with 17 free parameters. 

We perform an error analysis using the standard Hessian method introduced in \cite{Pumplin:2001ct}. We expand the $\chi^2$ function
around the minimum and define the Hessian matrix for our 17 free parameters $a_i$
\begin{equation}
	\chi^2(\{a\}) \approx \chi_0^2 + \frac12 \sum_{i,j} H_{ij}(a_i-a^0_i)(a_j-a^0_j) = 
	\chi_0^2 + \frac12 \sum_{i,j} \frac{\partial^2 \chi^2}{\partial a_i \partial a_j}\,(a_i-a^0_i)(a_j-a^0_j)\,.
\end{equation}
After diagonalization of the Hessian, we obtain the eigenvalues $\epsilon_k$ and eigenvectors $v_i^{(k)}$
\begin{equation}
	\sum_j H_{ij} v_j^{(k)} = \epsilon_k v_i^{(k)}\,,\qquad\quad z_k = \sqrt{\epsilon_k} \sum_j v_j^{(k)}(a_j-a^0_j)\,,
\end{equation}
and we introduce new re-scaled coordinates $z_k$ defined so that the Hessian transforms into a unit matrix.
Using these new coordinates, we define vectors $S_k^\pm$ which all have the same magnitude $\sqrt{\Delta\chi^2}$ and correspond to
a shift along the eigenvector direction $z_k$ in both directions. We then can use these vectors to generate the error PDFs as
\begin{equation}
	(\Delta X)_{\rm max}^\pm = \sqrt{\,\sum_k (X(S_k^\pm) - X(S_0))}\,,
\end{equation}
where we allow for different error in each direction resulting in asymmetric error bands.
\section{Results and Conclusions.}
\begin{figure}[htb]
  \includegraphics[height=0.97\textheight]{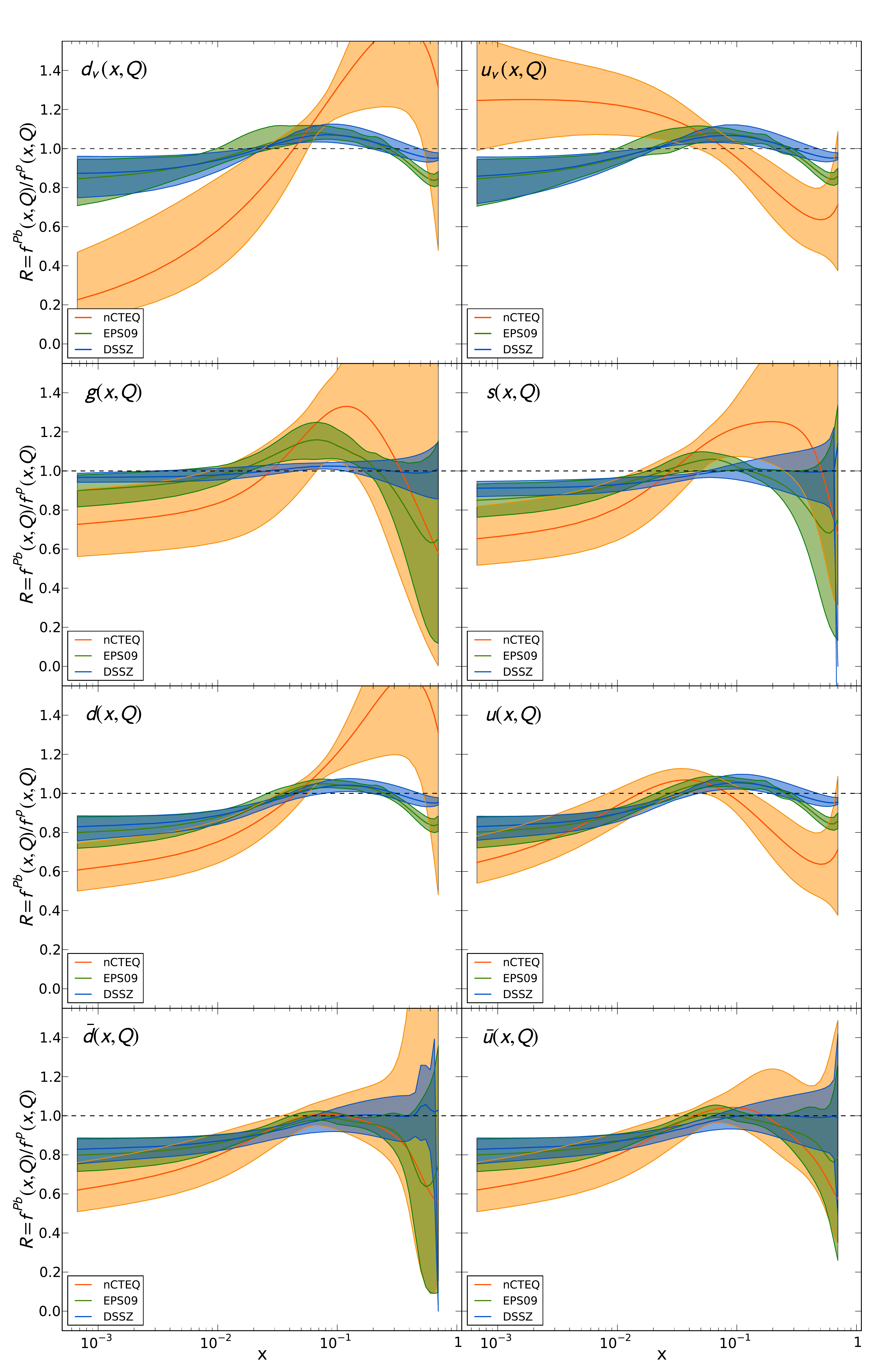}
\caption{We show nuclear modification ratios built using the PDFs themselves for lead (A=208) and at the scale $Q^2=100 {\rm GeV}^2$.}\label{Fig1}
\end{figure}

\begin{figure}[htb]
  \includegraphics[height=0.97\textheight]{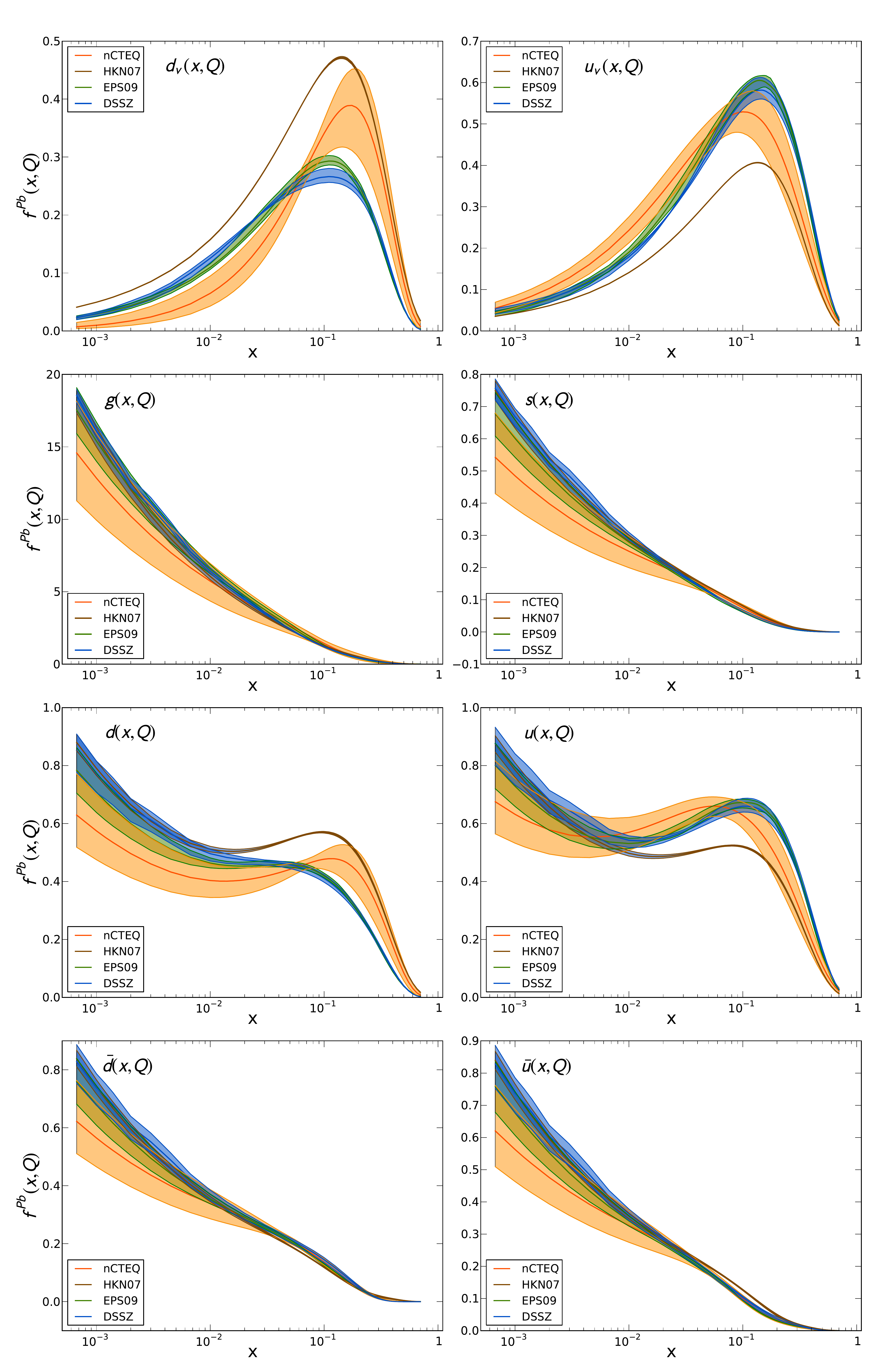}
\caption{We show nuclear parton distribution functions for lead (A=208) and at the scale $Q^2=100 {\rm GeV}^2$.}\label{Fig2}
\end{figure}

Results of our error analysis using the CTEQ fitting framework with 17 free parameters is shown in Figs.~\ref{Fig1} and \ref{Fig2}.
It is also compared to similar results by other groups \cite{Hirai:2007sx,Eskola:2009uj,deFlorian:2011fp}. In Fig.~\ref{Fig1}
we show the results in form of nuclear modification ratios which better highlight the nuclear effects with no reference to the 
underlying proton assumptions. In Fig.~\ref{Fig2} we show the true nuclear PDF with their uncertainties. In both Figures we observe 
that our error estimate is larger than the previous estimates which can also be attributed to larger number and a different choice 
of free parameters in our analysis.


\begin{thebibliography}{99}
\bibitem{Ball:2009mk}
R.~D.~Ball et~al.  [The NNPDF Collaboration],
{\em Nucl. Phys.},  B823:195--233, 2009, 0906.1958.
%
\bibitem{Martin:2009iq}
A.~D.~Martin, W.~J.~Stirling, R.~S.~Thorne and G.~Watt,
{\em Eur. Phys. J.}, C63:189--285, 2009, 0901.0002.
%
\bibitem{Nadolsky:2008zw}
Pavel~M. Nadolsky et~al.
\newblock {\em Phys. Rev.}, D78:013004, 2008, 0802.0007.
%
\bibitem{Owens:2012bv}
  J.~F.~Owens, A.~Accardi and W.~Melnitchouk,
  arXiv:1212.1702 [hep-ph].
%
\bibitem{Hirai:2007sx}
M.~Hirai, S.~Kumano and T.~H.~Nagai,
{\em Phys. Rev.}, C76:065207, 2007, 0709.3038.
%
\bibitem{Eskola:2009uj}
K.~J. Eskola, H.~Paukkunen, and C.~A. Salgado.
\newblock {\em JHEP}, 04:065, 2009, 0902.4154.
%
\bibitem{deFlorian:2011fp}
  D.~de Florian, R.~Sassot, P.~Zurita and M.~Stratmann,
  Phys.\ Rev.\ D {\bf 85} (2012) 074028.
%
\bibitem{Pumplin:2002vw}
J.~Pumplin et~al.
\newblock {\em JHEP}, 07:012, 2002, hep-ph/0201195.
%
\bibitem{Schienbein:2009kk}
  I.~Schienbein, J.~Y.~Yu, K.~Kova\v{r}\'{\i}k, C.~Keppel, J.~G.~Morf\'{\i}n, F.~Olness and J.~F.~Owens,
\newblock  {\em Phys. Rev.}, D80:094004, 2009, 0907.2357.
%
\bibitem{Kovarik:2010uv}
  K.~Kovarik, I.~Schienbein, F.~I.~Olness, J.~Y.~Yu, C.~Keppel, J.~G.~Morfin, J.~F.~Owens and T.~Stavreva,
  Phys.\ Rev.\ Lett.\  {\bf 106} (2011) 122301.
%
\bibitem{Pumplin:2001ct}
  J.~Pumplin, D.~Stump, R.~Brock, D.~Casey, J.~Huston, J.~Kalk, H.~L.~Lai and W.~K.~Tung,
  Phys.\ Rev.\ D {\bf 65} (2001) 014013.
\end{thebibliography}
\end{document}